\documentclass[conference]{IEEEtran}

\newcommand{\fakeparagraph}[1]{\vspace{.5mm}\noindent\textbf{#1.}}
\newcommand{\fakepar}[1]{\fakeparagraph{#1}}
\ifCLASSINFOpdf
  \usepackage[pdftex]{graphicx}
\else
\fi

\usepackage{xcolor}
\usepackage[cmex10]{amsmath}
\usepackage{amssymb}
\usepackage{caption}
\usepackage{subcaption}
\usepackage{url}

\usepackage[ruled,vlined,linesnumbered]{algorithm2e}

\newcommand{\MapName}{ELiD Map}

\begin{document}

\title{Elevated LiDAR based Sensing for 6G\\ - 3D Maps with cm Level Accuracy}
\author{\IEEEauthorblockN{Madhushanka Padmal\textsuperscript{1}, Dileepa Marasinghe, Vijitha Isuru,\\Nalin Jayaweera, Samad Ali, Nandana Rajatheva}
		\IEEEauthorblockA{\textit{Centre for Wireless Communications, University of Oulu,} Oulu, Finland \\
			\{dileepa.marasinghe, vijitha.isuru,
			nalin.jayaweera,
			samad.ali, nandana.rajatheva\}@oulu.fi}
			\textsuperscript{1}blog.padmal@gmail.com
	}
\maketitle

\begin{abstract}
%\boldmath

% Introduction to 6G enabler
One key vertical application that will be enabled by 6G is the automation of the processes with the increased use of robots. As a result, sensing and localization of the surrounding environment becomes a crucial factor for these robots to operate. Light detection and ranging (LiDAR) has emerged as an appropriate method of sensing due to its capability of generating detail-rich information with high accuracy. However, LiDARs are power hungry devices that generate a lot of data, and these characteristics limit their use as on-board sensors in robots. In this paper, we present a novel approach on the methodology of generating an enhanced 3D map with improved field-of-view using multiple LiDAR sensors. We utilize an inherent property of LiDAR point clouds; rings and data from the inertial measurement unit (IMU) embedded in the sensor for registration of the point clouds. The generated 3D map has an accuracy of 10 cm when compared to the real-world measurements. We also carry out the practical implementation of the proposed method using two LiDAR sensors. Furthermore, we develop an application to utilize the generated map where a robot navigates through the mapped environment with minimal support from  the sensors on-board. The LiDARs are fixed in the infrastructure at elevated positions. Thus this is applicable to vehicular and factory scenarios. Our results further validate the idea of using multiple elevated LiDARs as a part of the infrastructure for various applications.
%An elevated view of the ground plane will increase the field of view for a central entity to process path information. This enables coordination and control of multiple ground robots. 

 % add result and how this was done to abstract
 
%Track - 6ET – 6G Enabling technologies
\end{abstract}

% \fi
%\IEEEpeerreviewmaketitle

\section{Introduction}
%Main topics to be addressed in introduction:\\
%1. what do we expect in 6G - infrastructure based sensing\\
Innovation in wireless communications lead us towards the sixth-generation (6G) of communication networks that aim to deliver 1 Tbps peak data rate with 100 $\mu$s latency \cite{white_paper}. This opens up pathways to many vertical applications that can utilize high data rates and low latency links. Technologies such as mmWave and THz communications, massive MIMO  along with artificial intelligence have become key enablers for realizing such performances. One key challenge in utilizing these is to acquire precise positioning information. Positioning technologies such as global positioning system (GPS) are available for outdoor scenarios while indoor positioning is still a research challenge \cite{indoor-gps}. 

A promising vertical application of 6G is the factory automation where multiple robots perform their navigation and tasks with minimal human intervention \cite{factory-floor-robots}. Currently, these mobile robots  navigate utilizing on-board sensors such as cameras, proximity sensors and LiDARs. Moreover, these robots lack the perception of the whole environment as the on-board sensors have a limitation in both range and field-of-view (FoV). One solution to overcome this is to have inter-robot communications. However, sharing sensor information between mobile robots introduces a heavy burden on the communication links as those sensing technologies produce a huge amount of data which is in Gbps range. In our previous works, we proposed an infrastructure based sensing architecture with a coordinated set of elevated LiDARs (ELiDs) which develops a digital twin of the environment \cite{viima-project, viima-project2}. This architecture has the potential of delivering a solution to the positioning problem while overcoming the issue of limited perception of the environment to the robot.
%2. what is LIDAR and importance of LIDAR for factory automation.\\

LiDAR sensor is an active sensor which generates a point cloud using time of flight measurements between the emission of light and its return to the device. In contrast to cameras, LiDARs do not require pre-illumination of the environment and the point clouds generated are sparse compared to images. These features qualify LiDARs as sensors for automated navigation tasks in automobile industry and robotics. Merging the LiDAR streams from multiple LiDARs integrated to infrastructure as proposed, compensates for occlusion formed by obstacles, reduces sparsity and provides the potential of scaling up. Furthermore, it can be utilized to generate navigation commands for the mobile robots which can then operate with a minimal number of on-board sensors \cite{factory-automation}.
%\textcolor{blue}{This approach will mitigate the issue of local vision faced when using individual sensors of the robot. Path planning using proximity seeking characteristic is another approach to path planning and which is required the continuous vision of the robot when proceeding with the navigation process. 
%Most of the robotic applications use internal sensor data at each iteration of the navigation to avoid obstacles. But the global path planning method can help reduce the complexity that occurred due to the continuous searching process because the continuous seeking method will increase the complexity of memory and computation. Moreover, the power consumption of the sensors and robot will also increase due to proximity searching methods.}
%3. Importance of data fusion.\\
%4. Related works \\
%Why autonomous driving is a challenge in a large area like a factory. How 6G enablers would overcome these challenges and produce a method that can automate factory floors. How we are addressing this issue.


%Papers will be evaluated according to the following criteria:\\
%– Relevance and timeliness\\
%– Technical content and scientific rigour\\
%– Novelty and originality\\
%– Quality of presentation\\
%– Overall Recommendation

\fakepar{Contributions} In this paper we propose a novel approach to generate an \textbf{elevated LiDAR map (\MapName)} with cm level accuracy by merging multiple point clouds using their features and embedded sensors in the LiDAR. Enabled by the \MapName, we also present an application where navigation of a robot is carried out along the shortest path between two points in the mapped environment. %Ouster to LiDARs

The rest of the paper is organized as follows. In Section II, we describe the related work on the problem of merging point clouds and robot path planning based on point clouds. Next, we describe our proposed method of merging the point clouds to generate \MapName$\;$in Section III. In Section IV , the path planning based on the \MapName$\;$is presented. The results of the proposed method with comparison to real-world measurements and the navigation of the robot are presented in Section V. Finally Section VI concludes the paper.

\section{Related Work}
The problem of merging two point clouds has been in research for several years and multiple algorithms have been developed providing solutions as a result. Most of these methods are based on iterative closet point (ICP) algorithm to accurately register 3D point clouds. Even though ICP is a popular algorithm, there are a few limitations. It requires a proper initial value with accurate correspondence points \cite{3d-surface-reconstruction} and approximate registration between two point clouds to prevent it from falling into local extremes \cite{icp-geometric}. Attempting to estimate this registration automatically fails at instances when there are multiple similar views in point clouds \cite{auto-register-3D-point-clouds}. Instead, specific features such as curvature of objects is detected prior to estimating transformation to improve the iteration speed and accuracy in registering points using kd-trees \cite{icp-geometric,icp-kd-tree}. Even with such improvements, point clouds still need to be registered approximately.

When merging two point clouds, first step is to rotate and translate one point cloud while the other is kept fixed. Then both point clouds are concatenated yielding the merged point cloud. There are two main approaches in literature to generate merged point clouds which are as follows:

\subsubsection{Reference object based transformation}
\begin{figure}
     \centering
     \begin{subfigure}[b]{0.45\linewidth}
         \centering
         \includegraphics[width=\linewidth]{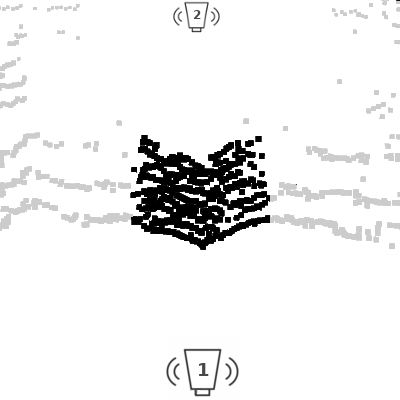}
         \caption{LiDAR 1 view}
         \label{fig:lidar-1-reference-box-view}
     \end{subfigure}
     \hfill
     \begin{subfigure}[b]{0.45\linewidth}
         \centering
         \includegraphics[width=\linewidth]{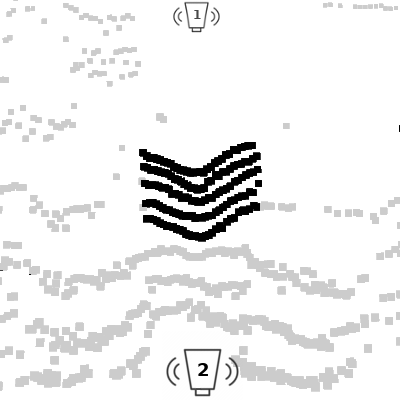}
         \caption{LiDAR 2 view}
         \label{fig:lidar-2-reference-box-view}
     \end{subfigure}
        \caption{Different view points from LiDAR sensors}\vspace{-2\baselineskip}
        \label{fig:different-view-points}
\end{figure}

While estimating the transformation matrix, restricting the view of a point cloud to a single object improves the transformation accuracy. Such an object is taken as a reference by multiple LiDAR sensors and a transformation is initially estimated. This transformation is then directly applied to the point cloud enclosing the object. This object should be rotation variant unlike cylindrical objects to help capture the correct view point \cite{3d-surface-reconstruction}. Multiple reference objects are used in literature not limited to cubes \cite{multi-lidar-multi-camera}, tetrahedrons \cite{lidar-camera-trihedron}, planes and wall boundaries \cite{3d-surface-reconstruction} to generate optimum results. It would require complete or partly manual intervention to identify where the object is placed inside the point cloud \cite{object-detection-in-pointclouds}. There are instances when cubes and tetrahedrons may fail being optimum reference objects due to the lack of knowledge of the surrounding environment. 
Figure \ref{fig:different-view-points}(\subref{fig:lidar-1-reference-box-view}) shows a point cloud segment captured from a LiDAR placed in front of a cubical. Figure \ref{fig:different-view-points}(\subref{fig:lidar-2-reference-box-view}) is from another LiDAR placed behind the same object and the points look more organized. Reference objects based transformation would disregard the 180$^{\circ}$ rotation of frames as they look identical. In such scenarios, cameras are used to gain a better insight about the surroundings.

\subsubsection{Camera based point cloud transformation}
Camera based methods utilize intrinsic parameters such as aspect ratio and extrinsic parameters such as relative position of the camera. World frame in a point cloud is identified using these parameters and a planar surface is estimated from LiDAR measurements\cite{3d-surface-reconstruction}. Cameras need to be calibrated to identify surfaces and powerful methods exist in image processing domain for that. Such methods process and merge images to use alongside LiDARs to estimate the transformation parameters \cite{multi-lidar-multi-camera}. Rodrigues et al. have developed a manual calibration method with a plane having circular holes. The reference simplicity reduces image noise, however, it does not guarantee an optimum estimate \cite{lidar-camera-1}. An automated solution was introduced by Alismail et al. where they use a center marked black circle and it can be used for calibrations during operation as well \cite{lidar-camera-2}. Object detection problem with a planar checkerboard pattern is used by Geiger et al. to identify extrinsic camera parameters for faster calibration. Still, it requires multiple references to localize in an environment \cite{lidar-camera-3}.

%Reference objects are equally used in both camera based \cite{multi-lidar-multi-camera} and LiDAR based localization methods \cite{3d-surface-reconstruction}. Figure \ref{fig:lidar-1-reference-box-view} shows a point cloud segment captured from an Ouster LiDAR placed in front of a cubical. Figure \ref{fig:lidar-2-reference-box-view} is from another LiDAR placed behind the same object and it is more organized compared to Figure \ref{fig:lidar-1-reference-box-view}. If reference objects based methods were to use, it would disregard the 180$^{\circ}$ rotation of a frame as they look identical. In such scenarios, it is difficult to estimate the pose without the help of external cameras. Reference object based methods require manual intervention or partly automated to identify where the object is placed inside the point cloud map \cite{object-detection-in-pointclouds}. But the complexity of such algorithms avoided them being used in this work as the parameter estimation is a one time task at the system setup.

 Point cloud aided navigation has been in research for many years. Zhang et al. have developed simultaneous localization and mapping (SLAM) based architecture to navigate a robot using on-board LiDAR sensors \cite{lidar-pathpplanning}. Extending the work on SLAM, Mahmood et al. describes a method using a velocity ramp to solve the drifting issue in robots moving in a straight line \cite{lidar-pathplanning2}. Vandapel et al. propose a laser detection and ranging (LADAR) based solution to navigate robots in unstructured cluttered environments \cite{LADAR-pathplanning}. Also, path planning based on voxelized point clouds to represent the environment in a 3D grid can be found in literature \cite{voxelized-pathplanning}.

\section{Proposed Solution}
We propose a two-step solution to generate an \MapName. The first step estimates rotation parameters of point clouds and the second step estimates translation parameters. The two parameter sets are combined to generate the transformation matrix.The process of building the point cloud map of the surrounding environment starts by collecting data frames from multiple LiDAR sensors mounted at different view points. Let the fixed point cloud chosen as reference be denoted by $\mathcal{S}$. Every other point cloud $\mathcal{M}_i\in\mathbb{R}^{K_{\mathcal{M}_i}\times4}$ for $i\in(1,\dots,N-1)$ where $N$ is the total number of input point clouds and $K_{\mathcal{M}_i}$ is the number of points in the point cloud $\mathcal{M}_i$ is rotated and translated with respect to the reference point cloud $\mathcal{S}$ to generate a transformed point cloud $\mathcal{M}_i'$. All $\mathcal{M}_i'$ point clouds are then concatenated with $\mathcal{S}$ to generate the \MapName  $\:\Upsilon$.

A point cloud $\mathcal{M}$ can be rotated and translated about the three major axes (x,y and z) by multiplying itself with a transformation matrix $\mathbf{T}\in\mathbb{R}^{4\times4}$ as $\mathcal{M}\mathbf{T}$. The $\mathbf{T}$ matrix is a combination of a rotation matrix $\mathbf{R}\in\mathbb{R}^{3\times3}$ and a translation vector $\mathbf{t}\in\mathbb{R}^{3\times1}$. They are combined with a $4\times4$ identity matrix, leaving the fourth row intact to generate the $\mathbf{T}$ matrix.

\subsection{Rotation parameters}
The first step is to estimate roll ($\phi$), pitch ($\theta$) and yaw ($\psi$) angles around x, y and z axes respectively. The roll ($\phi$) and pitch ($\theta$) estimations are based on inertial measurement unit (IMU) sensor readings embedded in the point cloud data frames. This differs from state of the art methods where they use a whole point cloud or a part of it to estimate rotation parameters using ICP algorithm based implementations \cite{auto-register-3D-point-clouds}.

\subsubsection{Estimating roll ($\phi$) and pitch ($\theta$) angles} Relative rotation around x and y axes of a point cloud $\mathcal{M}$ with respect to a reference point cloud $\mathcal{S}$, is estimated using IMU sensor readings. These sensors are inbuilt with LiDAR modules and they capture linear acceleration values along x, y and z axes keeping the horizontal plane of LiDAR sensor as the xy plane. Let the linear acceleration values along x, y and z axes of the LiDAR sensor generating the point cloud be $g_{x_{\mathcal{M}}}$, $g_{y_{\mathcal{M}}}$ and $g_{z_{\mathcal{M}}}$. Similarly, let the corresponding linear acceleration values of the LiDAR sensor generating the reference point cloud be $g_{x_\mathcal{S}}$, $g_{y_\mathcal{S}}$ and $g_{z_\mathcal{S}}$. The relative roll ($\phi$) and pitch ($\theta$) angles can be derived using,\vspace{-\baselineskip}

\begin{equation}\label{eq:relative-roll}
\phi = \tan^{-1}\left[
\frac{g_{y_{\mathcal{M}}}\sqrt{g_{x_{\mathcal{S}}}^2+g_{z_{\mathcal{S}}}^2} - g_{y_{\mathcal{S}}}\sqrt{g_{x_{\mathcal{M}}}^2+g_{z_{\mathcal{M}}}^2}}{g_{y_{\mathcal{M}}}g_{y_{\mathcal{S}}} + \sqrt{\left(g_{x_{\mathcal{M}}}^2+g_{z_{\mathcal{M}}}^2\right)\left(g_{x_{\mathcal{S}}}^2+g_{z_{\mathcal{S}}}^2\right)}}\right],
\end{equation} and\vspace{-\baselineskip}

\begin{equation}\label{eq:relative-pitch}
    \theta = \tan^{-1}\left[\frac{g_{x_{\mathcal{S}}}g_{z_{\mathcal{M}}}-g_{x_{\mathcal{M}}}g_{z_{\mathcal{S}}}}{g_{z_{\mathcal{M}}}g_{z_{\mathcal{S}}}+g_{x_{\mathcal{M}}}g_{x_{\mathcal{S}}}}\right].
\end{equation}

\begin{figure}
    \centering
    \includegraphics[width=\linewidth]{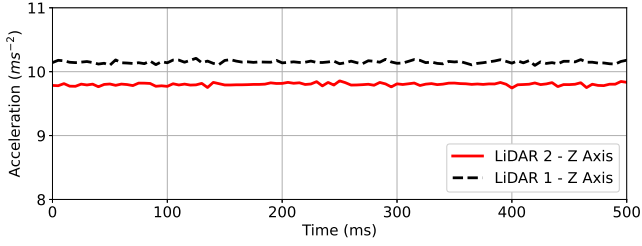}
    \caption{Variations and offsets in IMU readings.}\vspace{-2\baselineskip}
    \label{fig:imu-reading-offsets}
\end{figure}

This method cannot be directly implemented using LiDAR sensors out of the box. The reason is that, even when two such sensors are placed with the same orientation, they will not produce identical readings, as seen from Figure \ref{fig:imu-reading-offsets}. It requires a calibration step to nullify these offsets. In this calibration step, six sets of readings are taken by placing a LiDAR sensor facing towards and outwards from x, y and z axes. These readings consist of three sets of minimum and maximum linear acceleration values along each axis. Let the time average of these values along an arbitrary axis $\omega$ be $g_{\omega (\min)}$ and $g_{\omega (\max)}$. The corrected linear acceleration value $g_{\omega c}$ along the same axis for a reading $g_{\omega r}$ can then be interpreted as,\vspace{-\baselineskip}

\begin{equation}\label{eq:calibration-for-imu}
    g_{\omega c}=2\times9.80665\times\frac{g_{\omega r}-g_{\omega (\min)}}{g_{\omega (\max)} - g_{\omega (\min)}}-9.80665.
\end{equation}

The factor $9.80665$ in (\ref{eq:calibration-for-imu}) which is the acceleration due to gravity is extracted from LiDAR sensor datasheet in \cite{lidar-datasheet}. Instead of raw readings from IMU sensor, these calibrated values should be used in (\ref{eq:relative-roll}) and (\ref{eq:relative-pitch}). Even with the calibrated readings, use of a single calculated value is discouraged due to the varying nature of IMU readings following a normal distribution. Accordingly, the time average value of multiple readings will likely yield more accurate measurements.

\subsubsection{Estimating yaw $(\psi)$ angle} 
\begin{figure}
    \centering
    \includegraphics[width=\linewidth]{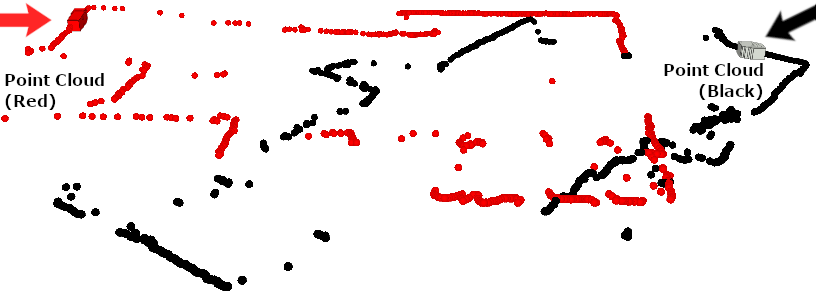}
    \caption{Selecting point segments to estimate relative yaw angle}\vspace{-\baselineskip}
    \label{fig:yaw-estimation}
\end{figure}

Relative rotation between two point clouds along z-axis is estimated using a feature in LiDAR point clouds. These point clouds consist of set of point rings. These rings can be interpreted as slices of the point cloud along z-axis. The first and last sets of point rings are highly warped. The middle set of point rings are almost parallel to the LiDAR xy-plane. This behavior of point rings prompts us to choose the middle point ring to estimate the relative yaw angle. First step in this process is to choose two point segments one from the middle point ring in each point cloud $\mathcal{M}$ (red) and the reference point cloud $\mathcal{S}$ (black) as shown by two arrows in Figure \ref{fig:yaw-estimation}.

When selecting these point segments, the user must consider choosing segments with at least three consecutive points such that the points approximately lie on a straight line capturing the same region of the environment represented in the point cloud. Colored blobs in Figure \ref{fig:yaw-estimation} satisfy these three requirements and the two point segments are then processed with random sample consensus (RANSAC) algorithm \cite{ransac-reference} to estimate a first order polynomial in the form of $f(x) = mx + c$ where $m$ is the gradient and which can also be represented as $\tan(\vartheta)$. Assume that for the selected set of points $M = \{{M}_i|{M}_i \in \mathcal{M},i = 1,\cdots,8\}$ the return value from RANSAC algorithm is its gradient $\tan(\vartheta_{\mathcal{M}})$. Similarly for a selected set of points $S=\{{S}_i|{S}_i \in \mathcal{S},i = 1,\cdots,8\}$ the return value is its gradient $\tan(\vartheta_{\mathcal{S}})$. Then the relative yaw ($\psi$) angle between the two point clouds can be calculated using\vspace{-\baselineskip}

\begin{equation}\label{eq-relative-yaw-angle}
    \psi = \tan^{-1}\left(\vartheta_{\mathcal{M}}\right) - \tan^{-1}\left(\vartheta_{\mathcal{S}}\right).
\end{equation}

Point cloud data points from a LiDAR sensor are always embedded with a noise component in the time of flight measurement readings. Hence the coordinate values of each point in the point cloud always vary along all three axes. This variance depends on few factors such as the type and material of the reflecting surface, Angle of Attack (AoA) on the reflective surface, distance from LiDAR sensor to the surface, timing errors in LiDAR sensor etc. Therefore obtaining an instantaneous value for the relative yaw angle using (\ref{eq-relative-yaw-angle}) by considering only a single instance of points segment is not guaranteed to produce a valid estimate for the required yaw angle. As a result, multiple readings are recorded on the same point segment selection to get multiple angle estimates and the average value is considered as the correct relative yaw angle.

The relative roll, pitch and yaw angles can then be used to populate the rotation matrix $\mathbf{R}$ as,
\begin{equation}\label{eq:rotation-matrix}\begin{bmatrix}
    \text{c}(\psi) & -\text{s}(\psi) & 0\\
    \text{s}(\psi) & \text{c}(\psi) & 0\\
    0 & 0 & 1
    \end{bmatrix}\begin{bmatrix}
    \text{c}(\theta) & 0 & \text{s}(\theta)\\
    0 & 1 & 0\\
    -\text{s}(\theta) & 0 & \text{c}(\theta)
    \end{bmatrix}
    \begin{bmatrix}
    1 & 0 & 0\\
    0 & \text{c}(\phi) & -\text{s}(\phi)\\
    0 & \text{s}(\phi) & \text{c}(\phi)
    \end{bmatrix},
\end{equation} where $\text{c}(.) = \cos(.)$ and $\text{s}(.) = \sin(.)$.

% \begin{figure}
%     \centering
%     \includegraphics[width=\linewidth]{ROS-VIIMA-System_Ext.png}
%     \caption{A highlevel view of the proposed ROS based system implementation.}\vspace{-\baselineskip}
%     \label{fig:ros-architecture}
% \end{figure}

\subsection{Translation parameters}
With the rotation matrix estimated, next step is to estimate the parameters for linear translation. This starts by rotating the point cloud using the estimated rotation matrix $\mathbf{R}$ in (\ref{eq:rotation-matrix}). Let us denote the rotated point cloud as an intermediate point cloud $\hat{\mathcal{M}}$. The two point clouds will now only have a linear offset between them. Estimating the linear translation parameters imply the same as estimating these offset values. This process is carried out along one axis and then continued with another axis until all three axes are covered. The implementation is derived from ICP algorithm. User will then select \textbf{two points} from a planar surface, one from the point cloud and the other from the reference point cloud. When selecting these points, the following requirements should be satisfied.

\begin{itemize}
    \setlength\itemsep{0pt}
    \setlength\parskip{0pt}
    \item Both points have to be on fairly flat surfaces facing each other along the selected axis.
    \item Surfaces on which the points are preferred to be from the same region in real environment.
    \item If they are not on the same region, the surfaces should belong to a similar planar region (eg: long running walls).
\end{itemize}

Elaborating more on these requirements, fairly flat surfaces are required to ensure that the neighboring points surrounding the selected point lie approximately on the same plane. The surfaces on which the points are located, should be the same surface. If not, they should at least belong to the same region in the environment to ensure that the offset is estimated against a common viewpoint. Such a scenario could occur when the two point clouds do not have an overlapping region in the real environment, but they capture two different segments of a single long wall. User intervention is required at this point to select such two points from the point clouds subject to the given constraints.

Once the two points are selected, four neighboring points closest to the selected point are filtered from each point cloud. When picking these four points, two of them are selected from the same point ring. The other two points picked from the two rings above and below. The filtered points in reference point cloud can be represented as $S^\omega = \{S_i|S_i\in\mathcal{S}, i=1,\cdots,5\}$. Similarly the filtered points from intermediate point cloud $\hat{\mathcal{M}}$ can be represented as $\hat{M}^{\omega} = \{\hat{M}_i|\hat{M}_i\in\hat{\mathcal{M}}, i=1,\cdots,5\}$.

The two sets of points $\hat{M}^{\omega}$ and $S^{\omega}$ are processed with the ICP algorithm to estimate the translation vector component along the axis of interest. Let this axis be $\omega$. Usually the ICP algorithm will return a complete transformation matrix. Disregarding the rotational components, the final column where the translation vector is located is extracted. From this translation vector, the value corresponds to the $\omega$-axis is extracted and it will be the offset of the intermediate point cloud $\hat{\mathcal{M}}$ relative to the reference point cloud $\mathcal{S}$ along $\omega$-axis.

A similar approach is carried out with the other two axes which completes the estimation of all the translation parameters to populate transformation matrix $\mathbf{T}$. The same process is independently applied to every other point cloud $\mathcal{M}_i$ to get a unique transformation matrix $\mathbf{T}_i$. Once all the transformation matrices are estimated, the \MapName $\;\Upsilon$ can be derived from
\begin{equation}\label{eq:point-cloud-merger}
    \Upsilon = \mathcal{M}_1\mathbf{T}_1 + \dots + \mathcal{M}_{N-1}\mathbf{T}_{N-1} + \mathcal{S}.
\end{equation}

\begin{figure}
    \centering
    \includegraphics[width=\linewidth]{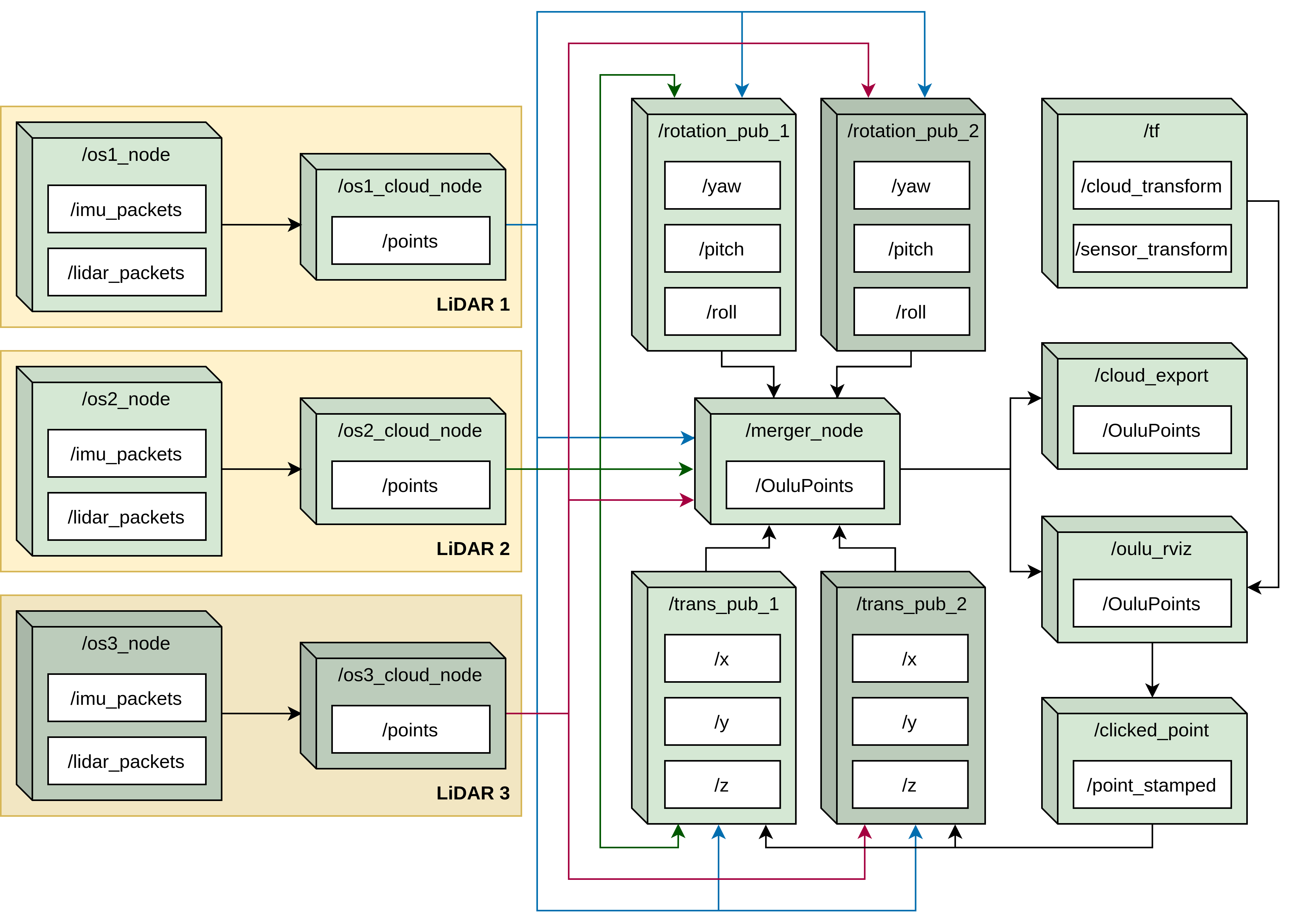}
    \caption{A highlevel view of the proposed ROS based system implementation.}\vspace{-\baselineskip}
    \label{fig:ros-architecture}
\end{figure}

This abstract method of generating a single point cloud combining multiple point clouds can be integrated into a practical implementation using robot operating system (ROS). Figure \ref{fig:ros-architecture} illustrates the high level view of the proposed system architecture to process, generate and export an \MapName$\;$using three different LiDAR sensors. Most commercially available LiDAR sensors provide an interface to integrate their sensors into ROS. ROS also provides a platform to process and visualize sensor data. This open source platform inherits faster execution of programs based on C++ programming language and support GPU based applications.

% \begin{figure}%[ht]
%     \centering
%     \includegraphics[width=\linewidth]{IEEEtran5/Application_EUCNC.jpg}
%     \caption{Overview of the path generation process.}\vspace{-\baselineskip}
%     \label{fig:ros_archi_proposed_sol}
% \end{figure}

\section{Robot path planning with \MapName} \label{Application}

In this section, we consider a factory automation scenario with the support of an \MapName. The map is used to plot the path and steer a ground robot moving towards its destination using a minimal number of on-board sensors mounted. The process begins by voxelizing the generated \MapName$\;$to reduce the resolution without losing the precision required for robot navigation  while avoiding obstacles. Then the voxelized map is used to estimate the shortest path between given two points on the \MapName$\;$ using an algorithm based on breadth first search (BFS) and Dijkstra's algorithm. 
\begin{figure}[ht]
    \centering
    \includegraphics[width=\linewidth]{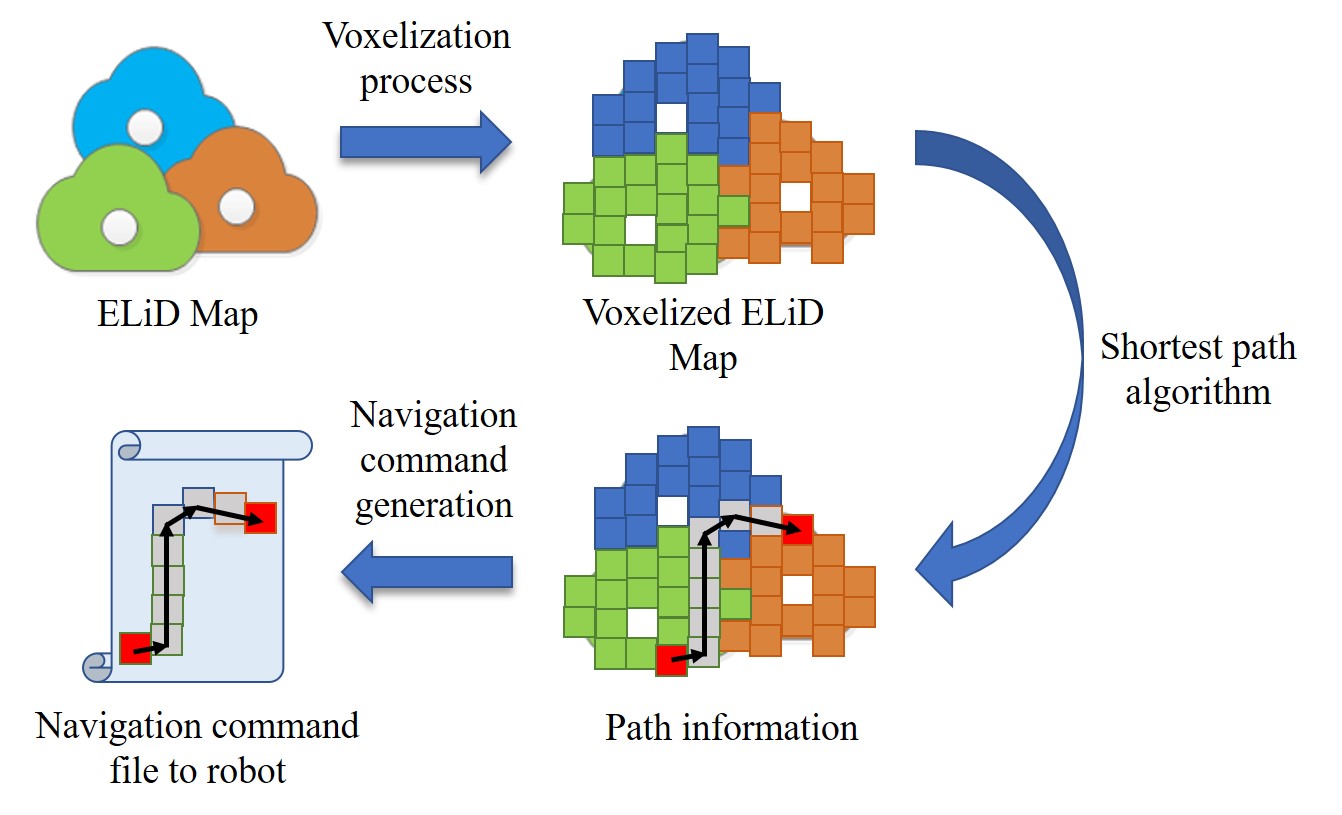}
    \caption{Overview of the path generation process.}\vspace{-\baselineskip}
    \label{fig:ros_archi_proposed_sol}
\end{figure}

\noindent  The estimated path is then translated into a set of basic navigation commands such as direction and move for a fixed time.  The offline computed commands are then uploaded to the robot over a wireless link. Figure \ref{fig:ros_archi_proposed_sol} illustrates the high level overview of implementing the path planning and navigation process.

\section{Results and Discussion}
\begin{figure}[h]
    \centering
    \includegraphics[trim={0mm 0 0mm 0},clip,width=0.85\linewidth]{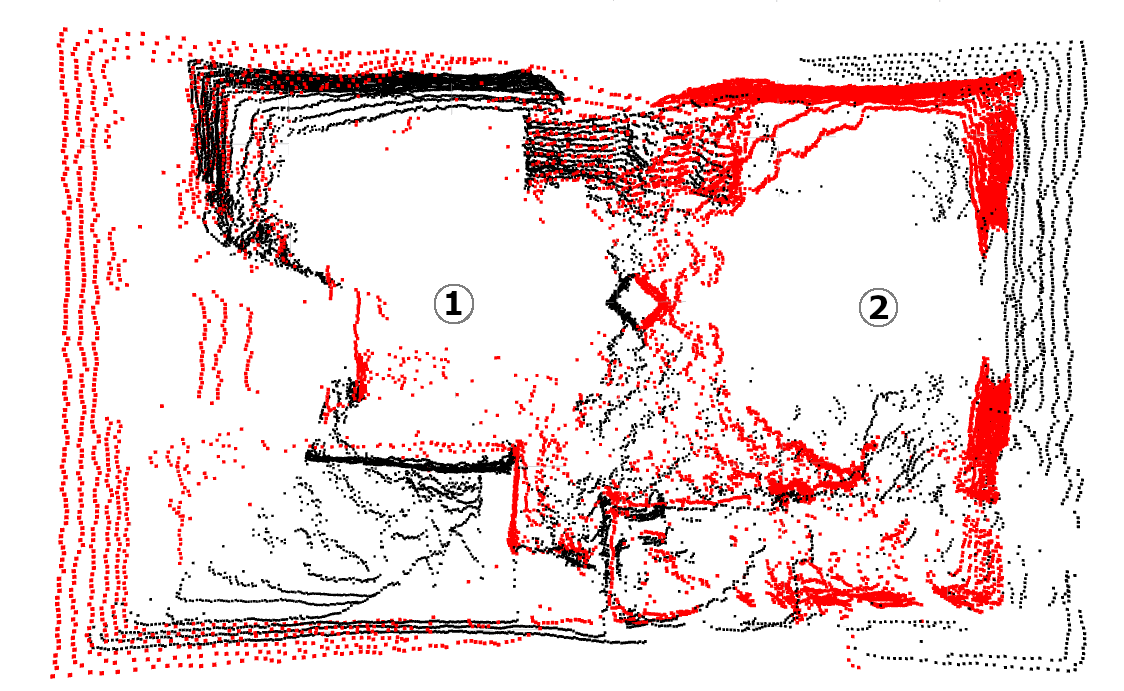}
    \caption{An \MapName $\;$using two point clouds (red and black).}%\vspace{-\baselineskip}
    \label{fig:merged-map-in-red-and-black}
\end{figure}

An \MapName$\;$of a room generated with two Ouster OS1-16 LiDAR sensors using the proposed solution is shown in Figure \ref{fig:merged-map-in-red-and-black}. The black colored point cloud is from a LiDAR placed at location 1 while the red colored point cloud is from a LiDAR placed at location 2. LiDARs are  positioned such that they have a linear offset along x-axis about two meters.

\subsection{Comparing merged point clouds with real layouts}
We are using ROS Rviz tool to visualize point clouds \cite{ros-rviz}. It allows measuring distances between points in real-time. A set of such measurements are tabulated in Table \ref{tab:realdata-vs-lidardata}. The actual length was measured using a measuring tape and the point length was measured in Rviz. The error column shows the difference between the two measurements and the error ranges between $4$ to $10$ cm regardless of the distance from the LiDAR sensors. Notice that this is a sufficient accuracy to utilize the \MapName$\:$for identifying the position of an object in the mapped environment which can be utilized in algorithms to improve wireless connectivity in mmwave/sub THz region.

% \begin{table}[t]
% \def\arraystretch{1.1}
% \begin{center}
%   \caption{Comparison between point cloud measurements and actual measurements}
%   \label{tab:realdata-vs-lidardata}
%   \begin{tabular}{l  l  l  l}
%     \hline
%     Scenario & Actual length (m) & Point length (m) & Error (m)\\
%     \hline
%     Minimum range & 0.70 & 0.7502 & 0.0502\\
%     Width of a cupboard & 0.93 & 0.9860 & 0.0560\\
%     Height of a cupboard & 0.92 & 0.9995 & 0.0795\\
%     Width of a room & 3.12 & 3.1689 & 0.0489\\
%     Length of a room & 5.00 & 5.0951 & 0.0951\\
%     \hline
%   \end{tabular}
%   %\vspace{\baselineskip}
% \end{center}
% \end{table}

\subsection{Time consumption}

\begin{figure}[ht]
    \centering
    \includegraphics[width=\linewidth]{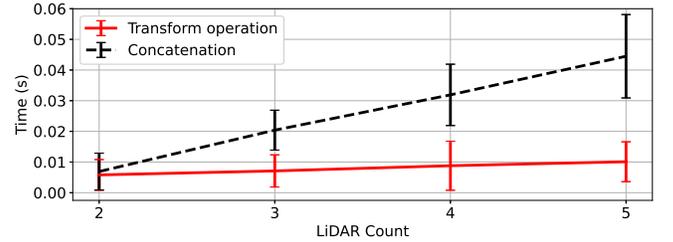}
    \caption{Time consumption for merge and transform multiple point clouds.}\vspace{-2\baselineskip}
    \label{fig:time-consumption-merge-transform}
\end{figure}

Timing analysis  was carried out combining two real-time LiDAR data streams and playback of recorded streams to resemble a multi-LiDAR setup to investigate the effect of increasing LiDARs to the computation time which is depicted in Figure \ref{fig:time-consumption-merge-transform}. The red curve shows time consumption for calculating transformation matrix and the black curve shows the time consumption for concatenating multiple point clouds. Timing values were calculated on a laptop running Ubuntu 18.04 LTS 64-bit operating system, with 2.3 GHz Intel CORE i5 processor and a 7.7 GB usable memory with no graphics processing unit (GPU) support. These timing values were collected over a period of 30 seconds for each LiDAR setup. Concatenation curve has a higher gradient since the accumulator has to handle an increasing number of points with an increasing number of LiDAR sensors. Transformation can be estimated independently causing a lower gradient.

A LiDAR sensor generates  a point cloud data stream at a rate of 20 Hz imposing a theoretical limit on the frequency of \MapName$\;$ generation. The proposed solution is able to process and transform point clouds from five different LiDAR sensors within 50 ms. This is within the point cloud generation interval of 50 ms. This timing value can be reduced using multiple threaded applications running in parallel with GPU support to merge point clouds in a binary tree pattern. Also, in a real implementation, such a multi-LiDAR system needs to be connected using Ethernet hubs to feed a continuous collision free LiDAR data stream to the central processor which will also add some delay.

%An optimized point cloud processor with GPU support would be able to manipulate a large number of LiDAR sensors generating point clouds.

\begin{table}[t]
\def\arraystretch{1.1}
\begin{center}
  \caption{Comparison between point cloud measurements and actual measurements}
  \label{tab:realdata-vs-lidardata}
  \begin{tabular}{l  l  l  l}
    \hline
    Scenario & Actual length (m) & Point length (m) & Error (m)\\
    \hline
    Minimum range & 0.70 & 0.7502 & 0.0502\\
    Width of a cupboard & 0.93 & 0.9860 & 0.0560\\
    Height of a cupboard & 0.92 & 0.9995 & 0.0795\\
    Width of a room & 3.12 & 3.1689 & 0.0489\\
    Length of a room & 5.00 & 5.0951 & 0.0951\\
    \hline
  \end{tabular}
  %\vspace{\baselineskip}
\end{center}
\end{table}

% \begin{figure}
%     \centering
%     \includegraphics[width=\linewidth]{time-consumption-lidar-merge-and-transform.eps}
%     \caption{Time consumption for merge and transform multiple point clouds.}\vspace{-\baselineskip}
%     \label{fig:time-consumption-merge-transform}
% \end{figure}

\subsection{LiDAR Setup time}
The proposed solution is implemented as a two step process. The first step is a one-time setup stage where the transformation matrix is estimated for each point cloud. The second stage is the operation stage where the concatenation takes place. Once the setup stage is completed, the LiDAR sensors are assumed to be fixed. The setup stage will require manual intervention and takes about twelve minutes to complete. This includes tasks such as calibrating IMU sensor readings (300 s), estimating rotation parameters (90 s) and translation parameters (300 s). This setup time can be improved further by reducing the manual intervention required to select specific features from the point clouds. For this, a  machine learning based feature detection algorithm can be implemented to choose such points automatically.

\begin{figure}[ht]
    \centering
    \includegraphics[width=0.85\linewidth]{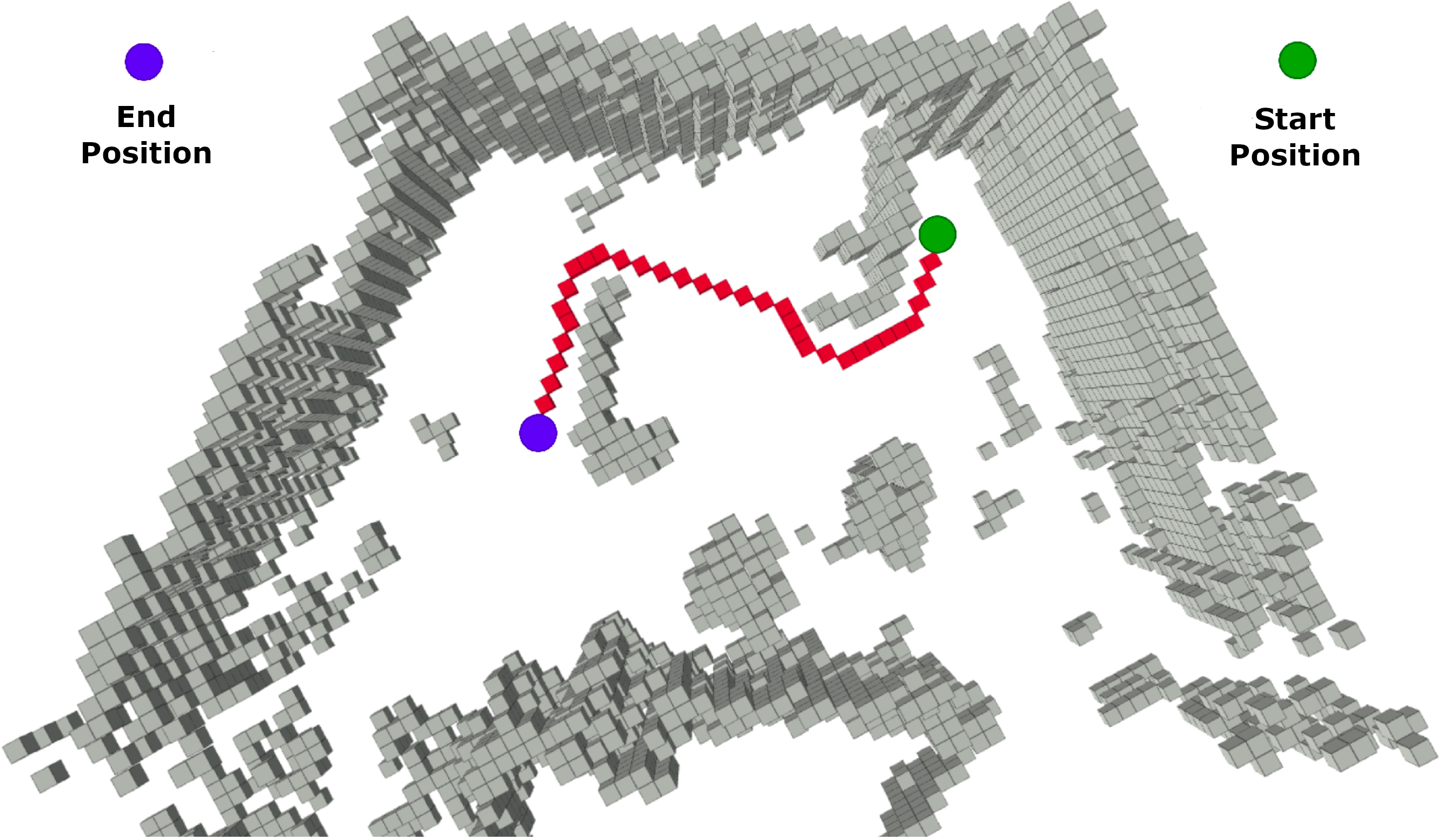}
    \caption{Planning shortest path in a point cloud}\vspace{-2\baselineskip}
    \label{fig:robot-path}
\end{figure}

\subsection{Path planning with voxelized \MapName}
\begin{figure}[ht]
    \centering
    \includegraphics[width=\linewidth]{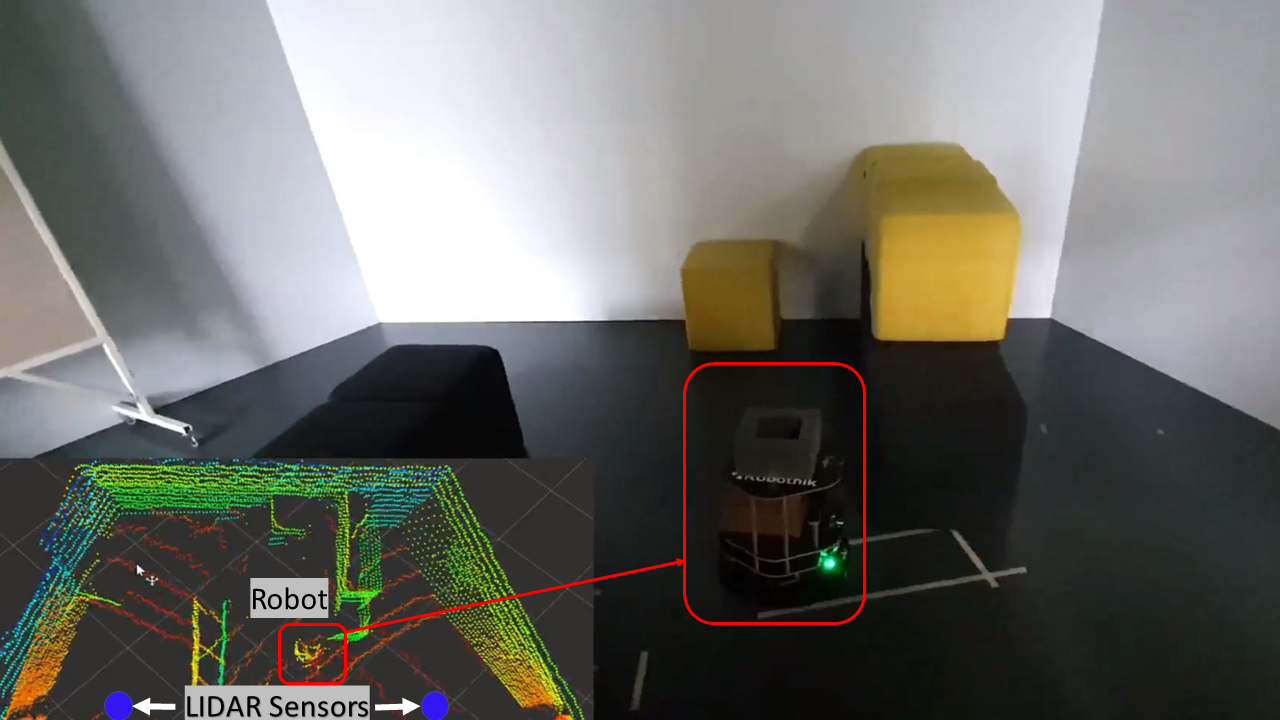}
    \caption{ An image of the setup where the robot is navigating along the path generated based on the \MapName }\vspace{-\baselineskip}
    \label{fig:realsetup}
\end{figure}

A ground setup with cubicles building up a simple maze was laid in a place in the university premises to demonstrate the application scenario. Two Ouster OS1-16 LiDAR sensors were placed on the two corners of the room and an \MapName$\;$was generated using the proposed method. A Kobuki TurtleBot was used as the ground robot. Figure \ref{fig:robot-path} shows the result of the calculated shortest voxel path in red between two points marked as green and blue dots on the voxelized \MapName. The offset between real world measurements and voxelized \MapName$\;$introduced as a result of the voxelization was taken into consideration while calculating the path. This would ensure that the robot will avoid any collisions with obstacles when passing them by. As described in Section \ref{Application}, the path is calculated offline and uploaded to the robot  using a Secure SHell (SSH) connection established over a WiFi link. Figure \ref{fig:realsetup} depicts an instance in the demonstration where the robot (highlighted in red) is utilizing the path information generated to reach the destination. The blue circles represent where the LiDAR sensors were placed.

\section{Conclusion}
In this paper, we have focused on a novel approach of generating an \MapName$\;$using integrated elevated sensors in the infrastructure targeting factory automation that plays a key role in enabling 6G evolution. First, we have generated the \MapName$\;$utilizing inherent properties of the LiDAR point clouds; the rings and the embedded IMU readings from the LiDAR sensors. Then, we have carried out experiments showing that the proposed method provides an \MapName$\;$with an accuracy of 10 cm when compared to real-world measurements. Finally, we have demonstrated the usability of the generated \MapName$\:$ with a practical implementation of path planning and navigation of a robot. In this work, path planning was done in an offline manner for a static environment. As an extension on the application, this can be improved where the central processor detects dynamic obstacles in near real-time and communicates the necessary navigation commands to avoid such obstacles to the robot through a fast wireless link.

% Realtime path updates
% accurate positioning can be 
% Wireless link between robots
% Realtime maps and navigation commands
% extend for moving objects in the environments.

% Communication aspect and the proposed solution is utilized in this path planning implementation.

% Digital twin is the point cloud map.


%\section*{Acknowledgment}




%\bibliographystyle{IEEEtran}
\bibliography{references}
\bibliographystyle{ieeetr}
%\begin{thebibliography}{1}


%\end{thebibliography}

\end{document}